\documentclass[prb,preprint,groupedaddress,showpacs]{revtex4}
\usepackage{graphicx} 
\begin{document} 
\bibliographystyle{apsrev}

\title{Electron bubbles in liquid helium: infrared-absorption spectrum}

\author{V\'{\i}ctor Grau}
\affiliation{Departament ECM, Facultat de F\'{\i}sica, Universitat de Barcelona. 
Diagonal 647, 08028 Barcelona, Spain}

\author{Manuel Barranco} 
\affiliation{Departament ECM, Facultat de F\'{\i}sica, Universitat de Barcelona. Diagonal 647,
08028 Barcelona, Spain}

\author{Ricardo Mayol}
\affiliation{Departament ECM, Facultat de F\'{\i}sica, Universitat de Barcelona. Diagonal 647,
08028 Barcelona, Spain} 

\author{Mart\'{\i} Pi} 
\affiliation{Departament ECM, Facultat de F\'{\i}sica, Universitat de Barcelona. Diagonal 647,
08028 Barcelona, Spain}

\date{\today}

\begin{abstract}

Within Density Functional Theory, we have calculated the energy of the
transitions from the ground state
to the first two excited states in the electron bubbles in liquid
helium at pressures from zero to about the solidification pressure.
  For $^4$He at low
temperatures, our results are in very good agreement with infrared absorption
experiments. Above a temperature of $\sim 2$ K, we overestimate the
energy
of the $1s-1p$ transition. We attribute this to the break down of the
Franck-Condon
principle due to the presence of helium vapor inside the bubble. Our results
indicate that the $1s-2p$ transition energies are sensitive not only to the size of the
electron bubble, but also to its surface thickness.
We also present results  for the infrared transitions in
the case of liquid $^3$He, for which we lack of experimental data.

\end{abstract}



\maketitle

\section{Introduction}

Excess electrons in liquid helium are known to form electron bubbles.
This is so because of the strong repulsion between the helium atoms,
which are very weakly polarizable, and the intruder electrons.
Experimental and theoretical spectroscopic
studies on electron bubbles have been carried out for many years
since the pioneering works of Northby and Sanders,\cite{Nor67}  and
of Dexter and coworkers.\cite{Fow68,Miy70}  Only some twenty years later,
Grimes and Adams were able to carried out a detailed analysis of the
infrared-absorption spectrum of the electron bubble in liquid $^4$He.
In a first work,\cite{Gri90}  they have used a photoconductive
mechanism to detect the transitions which operated only over
a limited region of the pressure-temperature ($P-T$) plane because
it appeared to be associated with trapping of bubbles on vortices
in the superfluid. Later on, they have improved their experimental 
apparatus and have observed the electronic transitions in direct
infrared absorption.\cite{Gri92}  This has allowed them to study the
spectrum of electron bubbles in a wider region of the $P-T$ plane
from $P=0$ to the solidification pressure, and up to temperatures
above 4 K instead of the maximum $T \sim 1.6$ K reached in photocurrent
experiments.

Recent experiments on cavitation in liquid helium have renewed the
interest in the study of single electron bubbles.\cite{Cla98,Kon03,Gri03}
Multielectron bubbles have also prompted some theoretical 
activity;\cite{Tem01,Tem03} while the stability of multielectron
bubbles is not yet clarified,\cite{Tem03,Mar03}
single electron bubbles are stable entities for pressures
above saturation pressure. Electron bubbles in helium droplets are
metastable objects whose lifetime has been measured
for both helium isotopes.\cite{Far98}

The simplest model to address electron bubbles in liquid helium
supposes that they are confined in a spherical well potential of
radius
$R$. The total energy of the electron-helium (e-He) system is then
written as a function of $R$
\begin{equation}
U(R)=E_e + 4 \pi R^2 \sigma +\frac{4}{3} \pi R^3 P \,\,\, ,
\label{eq1}
\end{equation}
where $E_e$ is the ground state  electronic energy,
$P$ is the pressure applied to the system, and
$\sigma$  is the surface tension of the liquid.
For  an infinite spherical well potential,
$E_e=\pi^2 \hbar^2 /(2 m_e R^2)$. This model is able to qualitatively
reproduce the experimental infrared-absorption energies, and is simple
enough to allow to calculate shape fluctuations of electron bubbles
and line shapes.\cite{Mar04}  It can be 
refined\cite{Fow68,Miy70,Gri90,Gri92}
by taking into account the finite depth of the well $V_0$,
which is about 1 eV, Ref. \onlinecite{Som64}. 
Once $U(R)$ has been minimized with respect to
$R$ and the radius of the equilibrium bubble $R_{eq}$ has been
determined, it is  
easy to obtain the energies of the $(n,l)$ excited states
and to compute the transition energies to the ground state if the
Franck-Condon principle holds,  i.e., if the absorption of a photon
by the electron bubble is a process much faster than the time needed
for the helium bubble to adapt itself to the electron
wave function in the excited state. 

Another key ingredient entering Eq. (\ref{eq1}) is the surface tension 
$\sigma$, which for $^4$He at zero $T$ and $P$ is  about
0.274 K \AA$^{-2}$, Ref. \onlinecite{Roc97}. This yields $R_{eq}=18.9$
\AA. The surface tension is only known along the liquid-vapor
coexistence line.
Consequently, to use Eq. (\ref{eq1}) one has to rely on model
calculations of $\sigma(P,T)$. The situation has been 
discussed in detail by Grimes and Adams, who concluded that in order to
perfectly fit their experimental results with
Eq. (\ref{eq1}), one has to take the depth of the well $V_0$ $P$-dependent
and a surface tension $\sigma$ independent of $P$, which seems difficult to
justify. For instance, early calculations\cite{Miy70,Ami66}
indicated that the surface tension $\sigma$ nearly doubles from $P=0$ to 25 atm.
For the sake of completeness, we present in the Appendix a calculation of
$\sigma(P)$ that also yields this increase, although we will not make
a substantial use of these results in our study.
It is worthwhile to stress that when both $V_0$ and $\sigma$ were made
$P$-dependent following the expected pressure dependences, the agreement 
between theory\cite{Miy70} and experiment was only qualitative, see Fig. 2 
of Ref. \onlinecite{Gri90}. For this reason, these authors concluded that more
sophisticated calculations of the properties of electron bubbles were
called for, incorporating the fact that the bubble density profile is
not abrupt but have a finite surface thickness.

In this work we present a theoretical description of infrared-absorption
$1s-1p$ and $1s-2p$ excitation energies of electron bubbles in liquid
helium. 
It belongs to the class of the more sophisticated calculations suggested by
Grimes and Adams, since it is not based on ad-hoc assumptions, but on
the use of a finite-temperature density functional (DF) approach
in conjunction with a realistic electron-helium  effective potential. 
DF methods have become increasingly popular in recent years as  useful
computational tools to study the properties of classical and quantum
inhomogeneous fluids,\cite{Eva89} especially for large systems for which these
methods provide a good compromise between accuracy and computational
cost, yielding results in agreement with
experiment or with more microscopic approaches. In the frame of DF theory,
the properties of an electron bubble approaching the surface of liquid
$^4$He have been studied by Ancilotto and Toigo\cite{Anc94} using
the so-called Orsay-Paris (OP) zero temperature finite-range 
DF\cite{Dup90} and the pseudopotential
proposed in Ref. \onlinecite{Kes65} as e-He interaction. 
This another key ingredient of the calculation, and
an accurate description of the infrared absorption can be only achieved
using an accurate e-He interaction. We want to mention that DF theory
has been succesfully applied to dynamical problems in bulk liquid
helium and helium droplets\cite{Elo02,Gia03,Leh04} using a
generalization of the OP density functional called Orsay-Trento (OT)
functional.\cite{Dal95}
Having no ad-hoc parameters to be adjusted in the calculation to describe
the infrared-absorption energies, we will show that the simplest form of
a DF for liquid helium, namely a zero range one, is not able to reproduce
the experimental data. As we are going to show, one must resort
to more sophisticated forms, such as the mentioned OP and OT finite range
functionals. This finding is an additional supporting evidence of the 
reliability of finite range functionals.

Density functional theory has also proven to be the most successful approach in
addressing cavitation in liquid helium so far.\cite{Xio89,Bar02,Bal02} It
incorporates in a self-consistent way the equation of state of bulk liquid
and surface tension of the liquid-gas interface as a function of temperature.
It allows for a flexible description of the electron bubble, incorporating
surface thickness effects. Within DF theory one avoids
the use of macroscopic concepts such as surface tension
and pressure at a nanoscopic scale;
however, it is a continuous, not an atomic description of the system.
Used in conjunction with a Hartree-type e-He potential,
we have recently shown that this approach quantitatively reproduces
the existing experimental data on cavitation of electron bubbles in
liquid helium below saturation pressure.\cite{Pi05} Consequently,
it is a tested framework to address other properties of electron
bubbles.

This paper is organized as follows. In Sec. II we briefly present
the density functional (DF) plus Hartree  electron-effective potential
approach we have employed, as well as some numerical details.
The structure of the electron bubble (radius and surface thickness) and
infrared-absorption energies are discussed in Sec. III, and
a Summary is presented in Sec. IV. Finally, we present in the Appendix
a model calculation of the surface tension as a function of pressure
that we have used to make some crude estimates of the electron bubble
radius.

\section{Density-functional approach and electron-helium interaction}

As in Ref. \onlinecite{Pi05}, our
starting point is the finite temperature zero-range DF of 
Ref. \onlinecite{Gui92}
that reproduces thermal properties of liquid $^4$He such as
the experimental isotherms
and the $^4$He liquid-gas coexistence line up to $T = 4.5$ K, and the
$T$ dependence of the surface tension of the liquid free surface.
We have taken the Hartree-type e-He effective potential
derived by Cheng et al.\cite{Che94} (see also Ref. \onlinecite{Ros95})
as e-He interaction.
This allows us to write the free energy of the system as
a functional of the $^4$He particle
density $\rho$, the excess electron wave function $\Psi$, and $T$:
\begin{equation}
F[\rho, \Psi, T] = \int d\vec{r}\, f(\rho, T)+
\frac{\hbar^2}{2\,m_e} \int d\vec{r}\, |\nabla \Psi(\vec{r}\,)|^2
+ \int d\vec{r}\, |\Psi(\vec{r}\,)|^2 V(\rho)  \; ,
\label{eq9}
\end{equation}
where $f(\rho, T)$ is the $^4$He free energy density per unit volume 
written as
\begin{equation}
f(\rho,T) = f_{vol}(\rho,T) + \beta \frac{(\nabla \rho)^2}{\rho}
+ \xi (\nabla \rho)^2  \; .
\label{eq10}
\end{equation}
In this expression,
$f_{vol}(\rho,T)$ consists of  the free energy density 
of a Bose gas, plus phenomenological density  and temperature
dependent terms written as
\begin{equation}
f_{int}(\rho, T)=
\frac{1}{2} b(T)\, \rho^2 + \frac{1}{2} c(T)\, \rho^{2+\gamma}
\label{eq9bis}
\end{equation}
that takes into
account the effective interaction of helium atoms in the bulk liquid.
The parameters of these terms and those of the density gradient terms
in Eq. (\ref{eq10}) have been adjusted so as to reproduce physical quantities
like the equation of state of the bulk liquid and the
surface tension of the liquid free surface.

The e-He interaction $V(\rho)$ is written as a function of the local
helium density \cite{Che94}
\begin{equation}
V(\rho) = \frac{\hbar^2 k^2_0}{2\,m_e}
+\frac{2 \pi \hbar^2}{m_e} \rho \,a_{\alpha}
- 2 \pi \alpha e^2 \left(\frac{4 \pi}{3}\right)^{1/3} \rho^{4/3} \; ,
\label{eq11}
\end{equation}
where $\alpha = 0.208\,$ \AA$^3$ is the static polarizability of
a $^4$He atom, and $k_0$ is determined from the helium local Wigner-Seitz
radius $r_s = (3/4 \pi \rho)^{1/3}$ by solving the trascendental
equation 
\begin{equation}
\tan[k_0(r_s-a_c)] = k_0\,r_s \; ,
\label{eq12}
\end{equation}
with $a_c$ and $a_{\alpha}$ being the scattering lengths arising from a
hard-core and from a polarization potential. 
We have taken \cite{Che94}
$a_{\alpha} = -0.06\,$ \AA, $a_c = 0.68\,$ \AA.

For given $P$ and $T$ values we have solved
Euler-Lagrange equations which result
from the variation of the constrained grand potential density 
$\tilde{\omega}(\rho, \Psi, T) = \omega(\rho, \Psi, T)
- \varepsilon |\Psi|^2 $, where 
\begin{equation}
\omega(\rho, \Psi, T) =  f(\rho, T)+
\frac{\hbar^2}{2\,m_e} |\nabla \Psi|^2
+ |\Psi|^2 V(\rho) - \mu \rho \; .
\label{eq13}
\end{equation}
It yields
\begin{equation}
\frac{\delta f}{\delta \rho} +|\Psi|^2\, \frac{\partial V}{\partial \rho}
= \mu
\label{eq14}
\end{equation}
\begin{equation}
-\frac{\hbar^2}{2\, m_e}\Delta \Psi + V(\rho) \Psi  = \varepsilon \Psi
  \; ,
\label{eq15}
\end{equation}
where $\varepsilon$ is the lowest eigenvalue of the Schr\"odinger
equation obeyed by the electron.
These equations are solved assuming spherical symmetry, imposing for $\rho$
that $\rho'(0) = 0$ and
$\rho(r \rightarrow \infty) = \rho_b$, where $\rho_b$ is the density
of the bulk liquid, and that the electron is in the $1s$ state.
Fixing $\rho_b$ and $T$ amounts to fix  $P$ and $T$, since the pressure 
can be obtained from the bulk equation of state
$P = - f_{vol}(\rho_b,T) + \mu \rho_b$, and
the $^4$He chemical
$\mu= \partial f_{vol}(\rho, T)/\partial \rho |_T$ is known
in advance. We have used
a multidimensional Newton-Raphson method\cite{Pre92}
for solving Eqs. (\ref{eq14}) and (\ref{eq15}),
after having discretized them
using $13$-point formulas for the $r$ derivatives. 
A fine mesh of step $\Delta r = 0.1\, $\AA$\,$
has been employed, and the equations have been integrated up to
$R_{\infty} = 150\,$ \AA $\,$ to make sure that the asymptotic bulk
liquid has been reached. After obtaining  the
equilibrium configuration ($1s$ state), the spectrum of the electron
bubble $\varepsilon_{nl}$ is calculated from Eq. (\ref{eq15}) keeping
frozen the helium density (Franck-Condon principle).

\section{Results}

\subsection{Liquid $^4$He}

Figure \ref{fig1} shows the $1s-1p$ transition energies (eV) as a function
of $P$ (atm) for $T =1.25$ K.
It can be see that the agreement between theory and
experiment is good from $P=0$ to the solidification pressure, and consequently,
the physical process seems well undestood.
However, a minor discrepancy appears at high pressures, and especially 
in the description of the $1s-2p$ transition
energies, as it can be seen in Fig. \ref{fig2}. This is not surprising,
since one would expect that the $2p$ state is more sensitive to fine
details of the bubble structure, in particular to its thickness,
because the $2p$  wave function penetrates deeper into the liquid.

We have improved the method of Sec. II to achieve a better agreement 
with experiment. The improvement is based on the observation that
zero-range DF's as the one described in Sec. II are fitted to reproduce the 
experimental surface tension of liquid helium at zero
temperature,\cite{Gui92,Str87}
and then, the $T$-dependence of $\sigma$,\cite{Gui92} and the thickness $t$
of the free surface -defined as the difference between the
distances at which the density equals 0.1$\rho_b$ and 0.9$\rho_b$,
where $\rho_b$ is the bulk density at the given $(P,T)$-
come out as predictions of the formalism. 
It turns out that at $T=0$, zero-range DF's overestimate $t$ by about 1 \AA.
Indeed, recent measurement of the $^4$He surface thickness yield values around 6 \AA,
Refs. \onlinecite{Har98,Pen00}, whereas the value predicted in 
Refs. \onlinecite{Gui92,Str87} is
about 7 \AA. In the case of $^3$He, the experimental value is about 
7.5 \AA,\cite{Har01} whereas the prediction using zero-range DF's is
about 8.5-9 \AA.\cite{Str87,Bar90}  We recall that
the surface thickness of liquid helium is a  quantity rather difficult to determine
experimental and theoretically,\cite{Edw78} and the dispersion of the values assigned
to it is large.\cite{Str87,Har98,Pen00,Bar90,Har01,Edw78,Osb89,Mar05}
Only recently, the mentioned values ($\sim$ 6-6.5 \AA $\,$ for $^4$He
and $\sim$ 7.5 \AA $\,$ for $^3$He) have emerged as likely accurate
determinations of the surface thickness of liquid helium free surface.

In the nineties, a new class of DF's has appeared that retains some of the simplicity of
the original zero-range DF's,  incorporating finite-range effects that are absolutely
necessary to address a wide class of physical phenomena, like elementary excitations in
bulk liquid and large inhomogeneities caused by the presence of impurities 
in bulk liquid and droplets, and also by substrates. Two such
DF's are the OP and OT functionals for $^4$He already commented.\cite{Dup90,Dal95}
It is remarkable
that they reproduce the experimental surface tension at $T=0$ without having imposed it
in their construction, and also the surface thickness. In particular, OP
yields $\sigma=0.277$ K \AA$^{-2}$, $t=5.8$ \AA, and OT yields $\sigma=0.272$ K
\AA$^{-2}$,  $t=6$ \AA.\cite{Dup90,Dal95}
A recent diffusion Monte Carlo calculation yields
$\sigma=0.281(3)$ K \AA$^{-2}$,  $t$=6.3(4) \AA.\cite{Mar05}

It is then quite natural to ask oneself if the remaining disagreement 
between theory and experiment shown in Fig. (\ref{fig1}) and especially in
Fig. (\ref{fig2}) might arise from a worse description of the electron bubble
surface in the case of zero-range DF's. Indeed, the error in the infrared
energies has been estimated\cite{Gri90} to be $\sim$ 1\%, smaller than
the difference between experiment and zero-range DF calculations.
To answer this question, we have repeated the calculations using the
zero temperature OT functional  -thermal effects on the excitation
energies are expected
to be small for $T =1.25$ K, and this is what we have found, see below-,
and also using a straightforward generalization at finite $T$ of the
OP functional, which is possible because of the similarities between 
OP and the  zero-range DF of Sec II.\cite{note} It is worth to know that
a generalization of the OT density functional at finite temperatures 
is also available.\cite{Anc00}

Density profiles of electron bubble at different
pressures and $T=1.25$ K, obtained with the OP and the zero-range functional,
are shown in Fig. \ref{fig3}, and in Fig. \ref{fig4} for the OT
functional at $T=0$ K. 
The excess electron squared wave functions $|\Psi|^2$ 
corresponding to the $1s, 1p,$ and $2p$ states are
also represented. For the zero-range DF we also give the value of the surface 
thickness $t$,
which goes from 6.1 \AA $\,$ at $P=0$ atm to 4.3 \AA $\,$ at $P=20$ atm. 
In the case of OP, and especially of OT, it is difficult to define 
the thickness, and even the bubble radius, because of the
density oscillations in the surface region 
(see also Ref. \onlinecite{Elo02}).
For reference, we indicate that the surface thickness of the electron
bubble
at $T=1.25$ K and $P=0$ atm obtained with OP is 5 \AA, one \AA $\,$
smaller that the value yielded by the zero-range DF.

Figures (\ref{fig1}) and (\ref{fig2}) also show the results obtained with OP and
OT. It can be seen that in this case, the agreement between theory and experiment 
is excellent.\cite{note1}
It is worthwhile to mention that Eloranta and Apkarian \cite{Elo02} have
also obtained the absorption energies up to $P \sim 12$ atm using the OT
functional and the pseudopotential of Jortner et al.\cite{Jor65} as e-He
interaction. Their results for the $1s-2p$ transition energies
are not in as good agreement with experiment as ours, likely because of
the different e-He interaction we have used.
Besides, the relevance of a correct description of the surface thickness
to reproduce the experimental data
has been overlooked in their work, which is mainly about the dynamics
of electron bubble expansion in liquid $^4$He.

Figure  (\ref{fig5}) shows the evolution of the radius of the electron bubble
with pressure determined by several authors using different 
methods,\cite{Gri90,Spr67,Ost73,Ell83}
and the value we have obtained using the zero-range and the OP and OT
functionals. For a given pressure, we have defined the radius of the bubble 
as the radial distance at which the helium density reaches the corresponding 
bulk value divided by two. It can be seen that our results
are compatible with the more recent experimental determinations, 
even those obtained with the
zero-range DF, which overestimate the radius of the electron bubble by
$\sim 0.5$ \AA.

We present in Fig. \ref{fig6} the variation of the $1s-1p$
excitation energy with $T$ at $P=2.9$ atm. Open circles are the
experimental data of Grimes and Adams, Ref. \onlinecite{Gri92},
and the dashed line is the
result obtained by these authors using their Spherical-square-well
(SSW) model
with the $T$ dependence of the surface tension as at saturation vapor
pressure.\cite{Iin85} Our results, displayed as filled circles,  have
been obtained using the finite temperature OP functional. It can be seen
that above $T=2$ K we overestimate the excitation energies.
In spite of the good agreement between experiment and the SSW model, we
believe it is coincidental, and that the failure of our method to
describe the experiment above $T=2$ K is the physical signature of the
breaking down of the Franck-Condon principle at high $T$.
Indeed, above $T\sim2$ K, it is no longer valid to assume
that the electron bubble is void of He atoms. It has been 
shown\cite{Cla98,Pi05}
that the bubble is progessively filled with helium vapor that must be
`dragged' by the electron in the course of its dipole excitation. This
dynamical effect has to be taken into account if one wants to have
a quatitative agreement with experiment above $T\sim 2$ K. Although it
is in principle possible to address this problem within time-dependent
density functional theory\cite{Gia03} using the finite
temperature OP functional or that of
Ref. \onlinecite{Anc00}, this is not a trivial issue and it is beyond the
scope of our work.

Finally, we have also calculated the cross-section of the transition
from the  $1s$ to the $1p$ and $2p$ excited states. For
photons of frequency $w$ polarized along the $z$ axis, the
cross section is given by
\begin{equation}
\sigma(w)= \frac{4 \pi^2 e^2}{c} w |\langle np | z | 1s\rangle|^2
\delta (E_{np}-E_{1s}-\hbar w)
\label{eqcross}
\end{equation}

We collect in Table \ref{table1}
the excitation energies (eV), oscillator strengths
and total cross sections (eV cm$^2$) to the first two $p$ states obtained
with the OT functional at $T=0$, $P= 0$ atm.
Our results are in full agreement with these of Ref. \cite{Fow68}.

\subsection{Liquid $^3$He}

We have also studied the infrared-absorption spectrum 
for electron bubbles in liquid $^3$He.
The application of DF theory to describe electron bubble explosions in
$^3$He proceeds as shown in Sec. II. We have used the zero-range DF
proposed in Ref. \onlinecite{Bar90} to describe the inhomogeneous
liquid, and also a finite-range DF, called from now on FR, 
obtained from the zero-range one following the procedure 
indicated in Ref. \onlinecite{Dup90} for the $^4$He case. The
e-He interaction is given again by
Eq. (\ref{eq11}) with the parameters corresponding to $^3$He, namely
$\alpha = 0.206\,$ \AA$^3$, and same values for $a_{\alpha}$  and
$a_c$.

We represent in Fig. \ref{fig7} the infrared-absorption 
energies for the $1s-1p$ and $1s-2p$ transitions.
It can be seen that the results are qualitatively similar
to those found for $^4$He, i.e., a fair insensitivity of
the $1s-1p$ transition energy to the detailed structure of the 
bubble surface, and an underestimation of the
$1s-2p$ transition energy by the zero-range DF.
The half density radius of the electron bubble $R_{1/2}$ 
goes from 22.5 \AA $\,$ at $P=0$ to 11.8 \AA $\,$ at $P=22.3$ atm.

The smaller excitation energies in the case of
$^3$He are due to the smaller surface tension for
this isotope, $\sigma = 0.113$ K \AA$^{-2}$,\cite{Iin85b}
which causes that
electron bubble radii are larger for $^3$He than for $^4$He 
(e.g., $R_{eq} =  23.5 \,$ \AA $\,$ at $P=0$ instead of 18.9 \AA),
yielding 
smaller excitation energies, as shown for instance by the simple model 
of Eq. (\ref{eq1}). In particular, at $P=0$ the model gives
for the ratio of the $1s-1p$ excitation energies the value
$(\sigma_3/\sigma_4)^{1/2}$, in good agreement with the results
obtained within the DF approach (see Tables).

Finally, we collect in Table \ref{table2}
the excitation energies (eV), oscillator strengths
and total cross sections (eV cm$^2$) to the first two $p$ states obtained
with the FR functional at $T=0$, $P= 0$ atm.
It can be seen that the
oscillator strengths and cross sections are sensibly the same for both
isotopes.

\section{Summary}

We have demonstrated the suitability of the density functional
approach to quantitatively address electron bubbles in
liquid He. 
On the one hand, we have shown that
the DF approach, in conjunction with a realistic electron-helium
interaction, is able to reproduce without any further assumption,
the low temperature infrared spectrum of electron bubbles 
experimentally determined by Grimes and Adams.\cite{Gri90,Gri92}
On the other hand, the method yields results that
agree with the experimental findings of Maris and co-workers
on cavitation of electron
bubbles below saturation pressure,\cite{Cla98,Su98}  in a
wide range of temperatures.\cite{Pi05}

We have shown that the analysis of infrared-absorption transitions of 
electron bubbles constitute a stringent test for the theoretical models
aiming a detailed description of the free surface of liquid helium, in
particular of its surface thickness.
This is at variance with electron bubble cavitation,
which seems to be sensitive only to global properties of the surface,
like its tension. Indeed, we have checked that the cavitation
pressure for electron bubbles in liquid $^4$He yielded by the
zero-range DF is -2.07 bar, whereas OP yields -2.13 bar,
and OT yields -2.08 bar.
These differences are far smaller than the experimental error bars, and
are partly due to the small differences in the surface tensions predicted by
these functionals.

There are some related problems that can be studied as a natural
extension of the work carried out until present within the DF frame.
In particular, the effect on the critical cavitation pressure 
of quantized vortices pinned to excess electrons.\cite{Cla98}
The infrared absorption
above $T \sim 2$ K is another open problem whose
quantitative description requires to relax
the Franck-Condon principle. Although they were not detected in the
experiments,\cite{Gri90,Gri92}  it would be also interesting to check
if there are other resonances in the infrared spectrum arising from
the state where the helium atoms are relaxed, even though this
state itself is not realized. For instance, in the case of electrons 
in crystals, this phenomenon, pertaining to polaron physics, is the 
origin of a peak in the electron optical spectrum
on top of the Franck-Condon spectrum, due to the resonance of this
virtual relaxed excited state with the applied radiation 
field.\cite{Dev72}
Both problems, presently unaffordable by any microscopic approach,
can be addressed within DFT and TDDFT. As these
are not trivial issues at all, they may be the next test grounds
to assess the capabilities and limitations of the DF theory applied to
liquid helium.

\appendix*
\section{}

In this Appendix we work out the zero temperature 
surface tension of bulk liquid $^4$He as a
function of $P$  for the zero range DF.
As we have previously stated, this is a model system, since
bulk helium can only be in
 equilibrium with its vapor along the coexistence line, and
thus for one single $P$ value at the given $T$.

We closely follow the method developed in Refs. 
\onlinecite{Gui92,Bar90}.
We take the planar interface perpendicular to the $z$ axis, and 
consequently, the liquid density is only function of $z$.
At given $P$, the equation of state determines the value of the bulk
density $\rho_b$ and chemical potential $\mu$. Starting
from Eqs. (\ref{eq10}) and (\ref{eq9bis}) taken at $T=0$, one writes
the Euler-Lagrange equation 
\begin{equation}
\frac{d f_{vol}}{d \rho} - \beta \left( \frac{\rho'^2}{\rho} +
2\,\frac{\rho''}{\rho} \right) - 2 \xi \rho'' = \mu  \; ,
\label{eqa1}
\end{equation}
where the prime denotes the $z$ derivative. Eq. (\ref{eqa1}) admits solutions that
go to $\rho_b$ when $z \rightarrow -\infty$, and
to zero when $z \rightarrow \infty$. For the Euler-Lagrange Eq. (\ref{eqa1}),
it is easy to obtain $\rho(z)$ -actually
$z(\rho)$- as indicated in Refs. \onlinecite{Gui92,Bar90}.
Density profiles corresponding to $P=0$, 5, 10, and 20 bar are shown
in Fig. \ref{fig8} as a function of $z$.
Note that for each profile, the location of the $z=0$ point 
is arbitrary, as it stems from the structure of Eq. (\ref{eqa1}), and
that the surface region is steeper the higher the pressure is.
The variation with pressure of the surface thickness is shown
in Fig. \ref{fig9}.

To obtain the surface tension, we first define a reference system in
which surface effects are not taken into account. It is a step-density
system of bulk density $\rho_b$ filling the $z<0$ halfspace,
whose free surface is arbitrarily located
at $z=0$. Placing -also arbitrarily- the overpressurized $^4$He density
profile in such a way that $\rho(z=0)=\rho_b/2$, we
obtain $\sigma(P)$ subtracting the grand potential of the overpressurized
system,
$\omega= f(\rho) - \mu \rho$, from that of the reference system,
$\omega_b= f_{vol}(\rho_b) - \mu \rho_b$, if $z\leq0$, and 0
anywhere else,
and dividing the difference by the surface of the planar interface.

Defining
\begin{eqnarray}
\Delta f & = & f_{vol}(\rho) - f_{vol}(\rho_b)
\nonumber\\
\Delta \rho & = & \rho - \rho_b 
\label{eqa2}
\end{eqnarray}
we have
\begin{eqnarray}
\sigma(P) & = & 2 \, \int_{-\infty}^0 (\Delta f - \mu \Delta \rho)\, d z
\nonumber\\
& + & 2 \, \int^{\infty}_0 
[f_{vol}(\rho) -\mu \rho + \beta \, \frac{\rho'^2}{\rho} +
 \xi \rho'^2]\, d z \; .
\label{eqa22}
\end{eqnarray}

Using that Eq. (\ref{eqa1}) can be integrated to yield \cite{Bar90}

\begin{equation}
\frac{d \rho}{d z} = - \left(\frac{\Delta f - \mu \Delta \rho}
{\frac{\beta}{\rho} + \xi}\right)^{1/2}
 \; ,
\label{eqa3}
\end{equation}
a close formula for the surface tension can be readily obtained:

\begin{eqnarray}
\sigma(P) = 2 \int^{\rho_b}_{\rho_b/2}
(\Delta f - \mu \Delta \rho)^{1/2}\, 
\left( \frac{\beta}{\rho} + \xi \right)^{1/2} d \rho+
\nonumber\\
\int_0^{\rho_b/2}
\frac{f_{vol}(\rho) - \mu \rho +\Delta f - \mu \Delta \rho}
{(\Delta f - \mu \Delta \rho)^{1/2}}\,
\left( \frac{\beta}{\rho} + \xi \right)^{1/2} d \rho
 \; .
\label{eqa4}
\end{eqnarray}
Note that the explicit knowledge of $\rho(z)$ is not needed to
determine $\sigma$. This is due to the particular form of
Eq. (\ref{eqa1}).\cite{Bar90}

Fig. \ref{fig9} shows the surface tension (K \AA$^{-2}$)
as as function of $P$ (atm). It can be seen that in the displayed
pressure range, $\sigma(P)$ increases almost linearly. The variation
we find from 0 to 20 atm is similar to that found in
Ref. \onlinecite{Ami66}.

\section*{Acknowledgments}
We would like to thank F. Ancilotto, E.S. Hern\'andez, and H.J. Maris
for useful
discussions. This work has been performed under grants FIS2005-01414
from DGI, Spain (FEDER), and 2005SGR00343 from  Generalitat de Catalunya.

\pagebreak

\begin{figure}[t]
\caption{ 
$^4$He $1s-1p$ transition energies (eV) as a function
of $P$ (atm) for $T =1.25$ K.
The open triangles are the observed 
points from the electron bubble photocurrent,\cite{Gri90}
and the open circles correspond to direct infrared absorption
measurements.\cite{Gri92} The dashed line represents the results obtained using
the zero-range DF discussed in Sec. II.
Filled circles connected with a dotted line are
the results obtained using the OP finite-range DF
at $T=1.25$ K, and filled diamonds connected with a solid line are
results obtained using the zero temperature OT finite-range DF.
}
\label{fig1}
\end{figure}

\begin{figure}[t]
\caption{ 
$^4$He $1s-2p$ transition energies (eV) as a function
of $P$ (atm) for $T =1.25$ K.
The open triangles are the observed 
points from the electron bubble photocurrent.\cite{Gri90}
The dashed line represents the results obtained using 
the zero-range DF discussed in Sec. II.
Filled circles connected with a dotted line are
the results obtained using the OP finite-range DF
at $T=1.25$ K, and filled diamonds connected with a solid line are
results obtained using the zero temperature OT finite-range DF.
}
\label{fig2}
\end{figure}

\begin{figure}
\caption{ 
Electron bubble density profiles in \AA$^{-3}$ (right scale) and
excess electron squared wave functions $|\Psi|^2$ in \AA$^{-3}$ 
(left scale), as a function of the radial distance $r$ (\AA)
for $T=1.25$ K and different pressures. 
The dashed lines correspond to the zero-range DF, and the
solid lines to the OP functional. 
The vertical thin lines indicate the
equilibrium radius $R_{eq}$ yielded by the simple electron bubble model
Eq. (\ref{eq1})
using the values of $\sigma(P)$ obtained in the Appendix. The value 
of the surface thickness $t$ is also given for the zero-range DF.
The electron squared wave functions $|\Psi|^2$ correspond to the OP 
calculation; dotted line, $1s$ state;
dot-dashed line, $1p$ state; thin solid line, $2p$ state.
}
\label{fig3}
\end{figure}

\begin{figure}
\caption{ 
Electron bubble density profiles in \AA$^{-3}$ (right scale) and
excess electron squared wave functions $|\Psi|^2$ in \AA$^{-3}$ 
(left scale), as a function of the radial distance $r$ (\AA)
for $T=0$ K and different pressures, obtained using the OT functional.
The electron squared wave functions $|\Psi|^2$ are represented as in 
Fig. \ref{fig3}, i.e., dotted line, $1s$ state;
dot-dashed line, $1p$ state; thin solid line, $2p$ state.
The vertical thin lines indicate the
equilibrium radius $R_{eq}$ yielded by the simple electron bubble model
Eq. (\ref{eq1})
using the values of $\sigma(P)$ obtained in the Appendix.
}
\label{fig4}
\end{figure}

\begin{figure}
\caption{ 
Radius of the electron bubble as a function of pressure.
Dashed line, zero-range DF results; filled circles(squares), OP(OT)
results at $T=1.25(0)$ K; dotted line, Ref. \onlinecite{Gri90} results.
Open squares, data with error bars, and open triangles are from
Refs. \onlinecite{Spr67,Ost73,Ell83}, respectively.
}
\label{fig5}
\end{figure}

\begin{figure}
\caption{ 
Variation of the $1s-1p$
excitation energy with $T$ at $P=2.9$ atm. Open circles are the
experimental data of Ref. \onlinecite{Gri92}. Dashed line, SSW model 
results.\cite{Gri92} Filled circles, finite temperature OP functional
results. The solid line has been drawn to guide
the eye.
}
\label{fig6}
\end{figure}

\begin{figure}
\caption{ 
$^3$He $1s-1p$ and $1s-2p$ transition energies (eV) as a function
of $P$ (atm) for $T=0$ K.
The dashed line is the result obtained using
the zero-range DF of Ref. \onlinecite{Bar90}.
Filled dots connected with a solid line are
the results obtained using the 
zero temperature finite-range DF.
}
\label{fig7}
\end{figure}

\begin{figure}
\caption{ 
Density profiles (\AA) for the overpressurized liquid $^4$He model system 
described in the
Appendix, as a function of $r$ (\AA) for several values of $P$ (bar). 
The results have been obtained using the zero range DF.
}
\label{fig8}
\end{figure}

\begin{figure}
\caption{ 
Surface tension (K \AA$^{-2}$, left scale) at zero temperature
as a function of $P$ (atm)
for the overpressurized liquid $^4$He model system described in the
Appendix. Also shown is the surface thickness $t$ (\AA, right scale).
The results have been obtained using the zero range DF.
}
\label{fig9}
\end{figure}

\newpage

\begin{table}
\caption{Transition energies, oscillator strengths and total
cross sections for the first two dipole excitations of electron
bubbles in $^4$He calculated with the OT functional 
at $T=0$ and $P = 0$ atm.
}

\begin{tabular}{|c|c|c|c|} \hline
Transition & Transition  & Oscillator & Total cross         \\
           & energy (eV) &  strength  & section (eV cm$^2$) \\ \hline
1s-1p      &   0.105     & 0.971    &  1.07 $\times 10^{-16}$ \\
1s-2p      &   0.488     & 0.0250   &  2.75 $ \times 10^{-18}$ \\ \hline 
\end{tabular}
\label{table1}
\end{table}

\begin{table}
\caption{Transition energies, oscillator strengths and total
cross sections for the first two dipole excitations of electron
bubbles in $^3$He calculated with the FR functional 
at $T=0$ and $P = 0$ atm.
}

\begin{tabular}{|c|c|c|c|} \hline
Transition & Transition  & Oscillator & Total cross         \\
           & energy (eV) &  strength  & section (eV cm$^2$) \\ \hline
1s-1p      &   0.071     & 0.972      &  1.07 $\times 10^{-16}$ \\
1s-2p      &   0.326     & 0.0248     &  2.73 $ \times 10^{-18}$ \\ \hline 
\end{tabular}
\label{table2}
\end{table}


\begin{thebibliography}{99}
\bibitem{note} This is achieved  making $T$-dependent the
characteristic distance $h$ at which the Lennard-Jones potential of OP
is screened,\cite{Dup90}  in such a way that at given $T$, the volume
integral of the screened potential equals the parameter $b(T)$ in
Eq. (\ref{eq9bis}), and taking the parameter $c$ of OP\cite{Dup90}
equal to the $c(T)$
parameter in Eq. (\ref{eq9bis}). These changes are complemented with 
the inclusion in the $T$ dependent  OP, of the free energy of the 
non-interacting Bose gas, as done in Ref. \onlinecite{Gui92}.
\bibitem{note1} To avoid overcrowding this figure, we have not
represented here the results obtained in Ref. \onlinecite{Gri90} using their
Spherical-square-well model, nor those obtained by the same authors
using the approach of Ref. \onlinecite{Miy70}. They can be seen in
Fig. 2 of Ref. \onlinecite{Gri90}.
\end{thebibliography}
\end{document}